\documentclass[conference]{IEEEtran}
\IEEEoverridecommandlockouts
\def\BibTeX{{\rm B\kern-.05em{\sc i\kern-.025em b}\kern-.08em
    T\kern-.1667em\lower.7ex\hbox{E}\kern-.125emX}}
\pdfoutput=1
\IEEEoverridecommandlockouts
\usepackage{cite}
\usepackage{amsmath,amssymb,amsfonts}
\usepackage{algorithmic}
\usepackage{graphicx}
\usepackage{textcomp}
\usepackage{xcolor}
\IEEEoverridecommandlockouts
\usepackage{cite}
\usepackage{amsmath,amssymb,amsfonts}
\usepackage{algorithmic}
\usepackage{graphicx}
\usepackage{textcomp}
\usepackage{url}
\ifCLASSINFOpdf
\else
\usepackage{caption}
\usepackage{subcaption}
\usepackage{subfig}
\fi
\usepackage{comment}
\usepackage{cite}
\usepackage{amsmath,amssymb,amsfonts}

\usepackage{mathtools}

\usepackage[utf8]{inputenc}
\usepackage{float}
\usepackage{physics}
\usepackage{multirow}
\usepackage{placeins}
\usepackage{amsmath,amssymb}

\usepackage{enumitem}

\newcommand{\paren}[1]{\left({#1}\right)}

\newcommand{\ith}[1]    {{#1}^{\underline{ \text{th}}}}

\begin{document}

\title{Performance of Large Aperture UCCA Arrays in a 5G User Dense Network}

\author{\IEEEauthorblockN{Md Imrul Hasan}
\IEEEauthorblockA{\textit{Electrical Engineering} \\
\textit{The University of Texas at Dallas}\\
Texas, USA \\
mdimrul.hasan@utdallas.edu}
\and
\IEEEauthorblockN{Mohammad Saquib}
\IEEEauthorblockA{\textit{Electrical Engineering} \\
\textit{The University of Texas at Dallas}\\
Texas, USA  \\
saquib@utdallas.edu}}

\maketitle
\begin{abstract}
\noindent   
The transmitted signals in the fifth generation (5G)  wireless networks suffer from significant path loss due to the use of higher frequencies in Sub-6 GHz and millimeter-wave (mmWave) bands. Inter-user interference in an ultra-dense network offers additional challenges to provide a high data rate. Therefore, it is desirable to generate narrow beams to extend the coverage of a 5G network by increasing antenna gain and improve its capacity by reducing the inter-user interference. This fact leads us to address the use of large aperture uniform concentric circular antenna (UCCA) arrays for 5G beamforming in massive multiple-input-multiple-output (MIMO) technology. Our analysis demonstrates that a UCCA with a larger antenna element spacing is capable of generating a significantly narrower beam with a moderate side-lobe level than a rectangular planar antenna (RPA) array while operating with the same number of antenna elements. This capability of the UCCA is analyzed to discover the performance gain of a 5G network.

\end{abstract}
\section{Introduction} 
In recent years, with the advent of the internet of things (IoT), and the increasing data rate of media-rich applications, the demand for wireless bandwidth has been expanding very rapidly. The fifth-generation (5G) network is introduced to support this high demand for spectrum \cite{af}. This network uses higher frequencies in Sub-6 GHz (S-band, C-Band, etc.) and millimeter-wave (mmWave) bands, both of which are prone to high path loss. Moreover, in a case when two user equipments (UEs) are spatially close to each other, their data rate decreases, and in the worst scenario, either of the two UEs can be dropped by the scheduler. A possible solution to these limitations is the use of massive multiple-input multiple-output (mMIMO) technology which offers higher gain and interference suppression capability by allowing its antenna array to focus narrow beams towards a user. These narrow beams also allow spatial multiplexing while increasing the number of UEs for the same time/frequency resources. Therefore, an antenna array capable of offering narrower beams is always desirable. 

There exist many types of geometries for 2-D mMIMO antenna arrays. Most of the works in the literature focus on the use of rectangular planar antenna (RPA) arrays \cite{rpa,rr2}. However, uniform concentric circular antenna (UCCA) arrays also have multiple advantages; the flexibility in array pattern synthesis and design \cite{a03}. The large aperture antenna array is introduced in order to decrease the correlation among the closely spaced UEs \cite{pratschner2019large}. Our analysis reveals that for a given number of antenna elements, an RPA array with a large aperture (element spacing is greater than or equal to half of the wavelength) results in many higher side lobes, whereas, a UCCA offers a significantly narrower main beam with moderate side lobe level (SLL). This observation leads us to investigate the performance of a large aperture UCCA for different element spacing in the context of 5G beamforming.

\section{Antenna Array Model}
\label{model}
\begin{figure}[t!]
\centering
\includegraphics[scale=0.55]{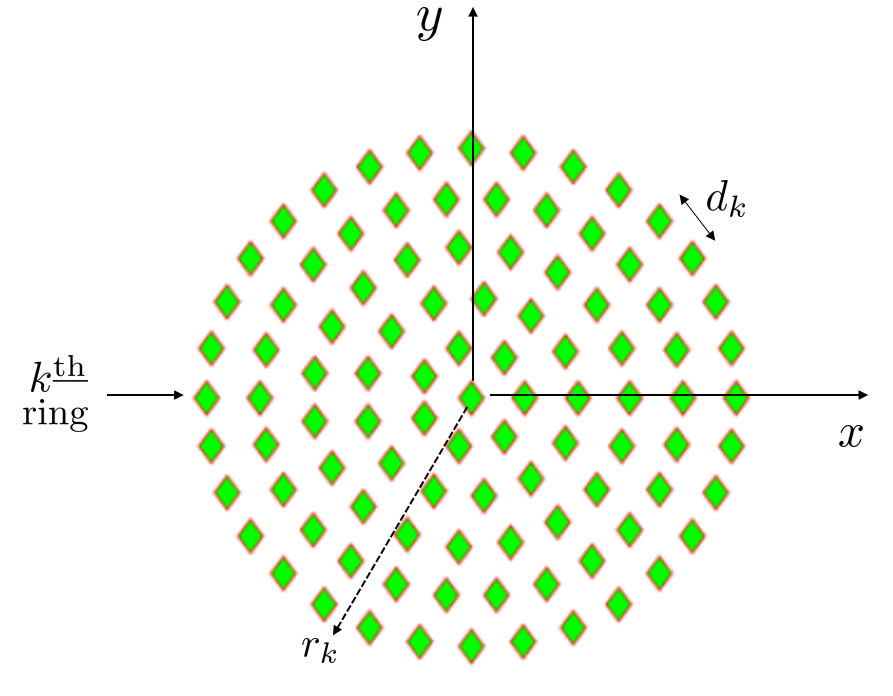}
 \caption {UCCA array geometry.}
 \label{Fig2}
\end{figure}

\begin{figure}[t!]
\centering
    \includegraphics[scale=0.81]{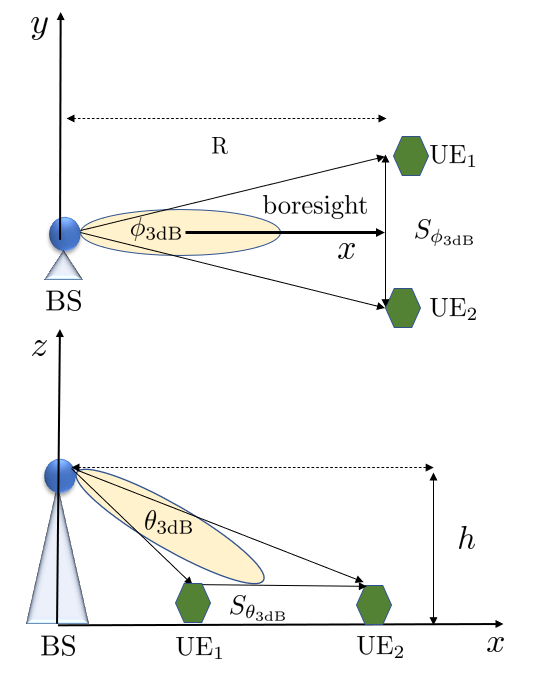}
    \caption{Inter-UE spatial separation in 2D.}
    \label{Fig3}
\end{figure} 

 Let's provide the mathematical model of a UCCA array assuming the desired signal of wavelength $\lambda$ impinges upon the array from the elevation angle $\theta_0$ and the azimuthal angle $\phi_0$. The geometry of an UCCA array is depicted in Fig. \ref{Fig2}, where the $\ith{i}$ ring in the array with a radius $r_i$, contains $N_i$ omni-directional elements; $i = 1, 2, ... , k$. If $d_i$ denotes the inter-element distance in the $\ith{i}$ ring, then $d_i=\Gamma \paren{2\pi r_i/N_i}$, where the function $\Gamma$ rounds up the fraction to the upper integer value. The array factor of the UCCA directed towards $(\theta_0,\phi_0)$ is given by \cite{b2}
\begin{equation}
    F_{\rm{UCCA}}(\theta,\phi)=1+\sum_{i=1}^{k}\sum_{j=1}^{N_i}B_{ij}e^{i(2\pi r_i/\lambda) f_{ij}(\theta,\, \phi)},
    \label{CCA}
\end{equation}
 where $B_{ij}$ the beamformer weight for the $\ith{j}$ element in the $\ith{i}$ ring, $$f_{ij}(\theta,\, \phi)=  \sin \theta  \cos (\phi- \phi_{ij})-\sin \theta _0 \cos (\phi_0- \phi_{ij})\,,$$ and $\phi_{ij}=2\pi (j-1)/N_i$.

\section{Beamforming Performance Criteria}
In 5G networks, the narrow beam formed by an mMIMO facilitates more UEs within the same spatial dimension. To evaluate how densely we can allow beams to support neighboring UEs in 3D, the idea of beam packing is explored considering an arbitrary sphere. This number is calculated by the maximum number of supported half-power beamwidths (HPBWs) in the elevation ($\theta_{\rm{3dB}}$), and the azimuthal ($\phi_{\rm{3dB}}$) planes. The overall beam packing gain can be defined as

\begin{equation}
    G_{\rm{BP}}=\paren{\frac{90^{\circ}}{\theta_{\rm{3dB}}}}\paren{\frac{360^{\circ}}{\phi_{\rm{3dB}}}}.
    \label{BP}
\end{equation}
Next, we relate the angular HPBW separation to UE separation distances. As shown in Fig. \ref{Fig3}, these distances can be approximated as \cite{b5}
\begin{equation}
    S_{\phi_{\rm{3dB}}}=2R \,\tan\{\phi_{\rm{3dB}}/2\}\,,
    \label{ds1}
\end{equation}
and 
\begin{equation}
   S_{\theta_{\rm{3dB}}}=R-h \, \tan\{\arctan(R/h)-\theta_{\rm{3dB}}\},
      \label{ds2}
\end{equation}

 \begin{figure}[t!]
\begin{minipage}[b]{0.45\linewidth}
\centering
\includegraphics[height=4.2cm, width=4.7cm]{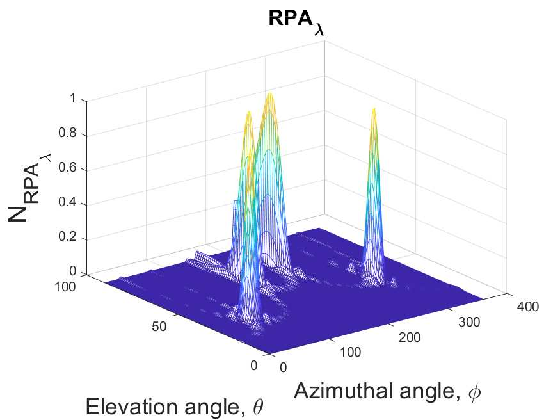}
\end{minipage}
\hspace{0.5cm}
\begin{minipage}[b]{0.47\linewidth}
\centering
\includegraphics[height=4.2cm, width=4.7cm]{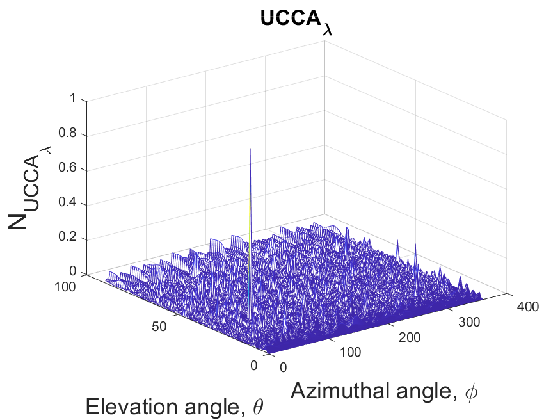}
\end{minipage}
\caption{Beam pattern for element spacing = $\lambda$.}
\label{a}
\end{figure}
\begin{figure}[t!]
\begin{minipage}[b]{0.45\linewidth}
\centering
\includegraphics[height=4.2cm, width=4.7cm]{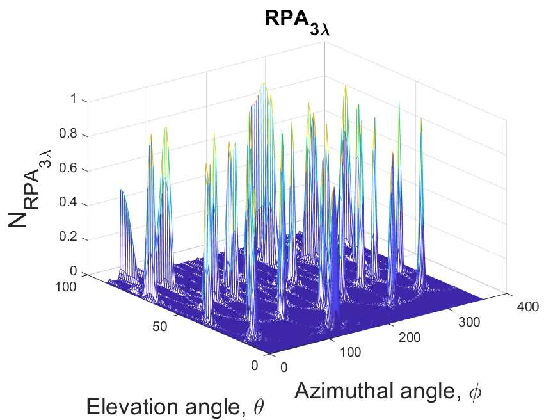}
\end{minipage}
\hspace{0.5cm}
\begin{minipage}[b]{0.47\linewidth}
\centering
\includegraphics[height=4.2cm, width=4.7cm]{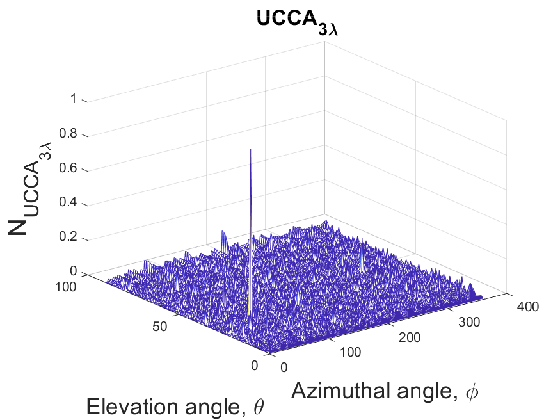}
\end{minipage}
\caption{Beam pattern for element spacing = $3\lambda$.}
\label{b}
\end{figure}
where $R$ is the range and $h$ is the antenna height. In (\ref{ds1}), $S_{\phi_{\rm{3dB}}}$ is the minimum inter UE distance at the boundary on the azimuthal plane. Similarly, in (\ref{ds2}) $S_{\theta_{\rm{3dB}}}$ denotes the minimum radial separation distance of two UEs on the elevation plane. Note that all these separation schemes prohibit half-power beam overlapping while suppressing the interference from any undesired direction.

Now, in order to analyze the quality of signals in 5G networks, we evaluate the SINR which is given by \cite{b3}
\begin{equation}
    {\mathrm  {SINR}}={\frac  {P}{I+N}}\,,
    \label{sinr}
\end{equation}
 where $P$ is the received power of the signal of interest, $I$ is the interference power of the other (interfering) signals in the network, and $N$ is the average power of the background noise. In general, the desired signal, interference, and noise signals are mutually statistically independent. The above SINR is directly related to the spectral efficiency, which can be written as \cite{pratschner2019large}
\begin{equation}
    \mathrm{SE}= \log_2\paren{1+\rm{SINR}}\\ \mbox{bps/Hz}.
     \label{se}
\end{equation}

\section{Numerical Study}


In this section, our objectives are a) to perform a comparative study of large aperture UCCA arrays with different element spacings on the system level using (\ref{BP})-(\ref{ds2}), and b) to provide insights into the link level performance of the arrays using (\ref{sinr}) and (\ref{se}). In this study, we use a uniform RPA array as the baseline performer. The beam pattern of the RPA array towards the desired user is well known and obtained by \cite{b1}
\begin{figure}[t!]

\centering
\includegraphics[scale=0.57]{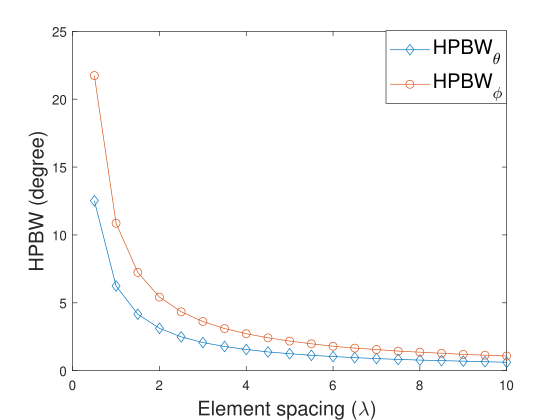}
\caption{HPBW for variable element spacing in a UCCA array.}
\label{c}
\centering
\includegraphics[scale=0.57]{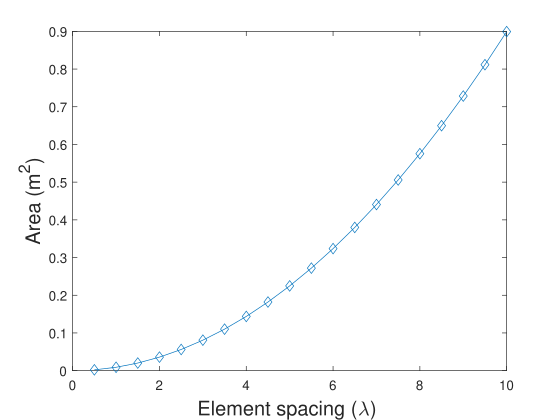}
\caption{Area size of a UCCA array with variable element spacing.}
\label{d}
\end{figure}
\begin{equation}
\centering
    F_{\rm{RPA}}(\theta,\phi)=\sum_{m=0}^{N_x-1}\sum_{n=0}^{N_y-1}A_{mn}e^{i(2\pi/\lambda)\{m \times h(\theta,\, \phi)+n\times g(\theta,\, \phi)\}},
    \label{UPA}
\end{equation}
where the number of elements along the $x$ and $y$ axis are $N_x$, and $N_y$ with element spacing $d_x$, and $d_y$.
Here $A_{mn}$ denotes the beamformer weight, and 
\begin{equation*}
    h(\theta,\, \phi) = d_x\paren{\sin\theta\cos\phi - \sin\theta_0\,cos\,\phi_0}\,,
    \end{equation*}
    \begin{equation*}
    g(\theta,\, \phi) = d_y\paren{ \sin\theta\sin\phi- \sin\theta_0\sin\phi_0}\,.
\end{equation*}

Assume, the frequency of interest is 28 GHz band, and the signal of interest is at $(\theta_0,\phi_0)=(30^{\circ},60^{\circ})$. For the UCCA arrays, we use the total number of rings, $k=5$, and equal inter ring and inter element spacings. These yield the number of antenna elements in the UCCA array, $N_{\rm{UCCA}}=98$. As for the RPA array beamforming, we use $N_x=N_y=10$, hence, the total number of elements, $N_{\rm{RPA}}=100$. The normalized beam pattern of the above arrays can be defined by
  \begin{equation}
      N_{\rm{array}}(\theta,\phi)=\paren{ |F_{\rm{array}}(\theta,\phi)|^2/|F_{\rm{array}}(\theta_0,\phi_0)|^2 }.
  \end{equation}
The beam patterns are depicted in Fig. \ref{a} and \ref{b} for RPA and UCCA arrays considering element spacing $>\lambda/2$ , Here, we have assumed uniform illumination ($A_{mn}=B_{ij}=1$) for both the array structures \cite{b1}.
 \begin{figure}[t!]
\centering
\includegraphics[scale=0.57]{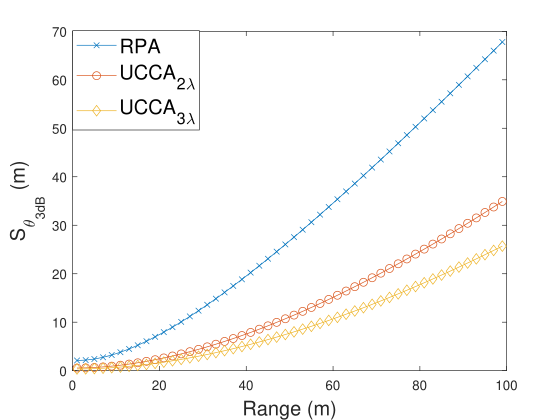}
\caption{Spatial separation distance on the elevation plane.}
\label{Fig12}

\centering
\includegraphics[scale=0.57]{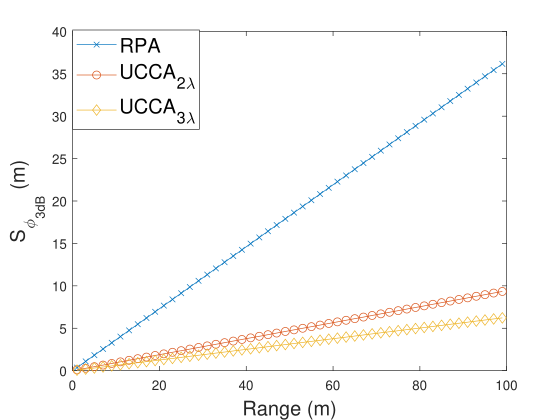}
\caption{Spatial separation distance on the azimuthal plane.}
\label{Fig121}

\end{figure}
Notice that although both the arrays have the same maximum gain towards the desired UE, the RPA array with element spacing  $>\lambda/2$ is highly vulnerable to interference from other directions. Therefore, a larger element spacing is not recommendable for an RPA array. On the other hand, the UCCA array is capable of producing significantly narrower beams without noticeably increasing the SLL as the element spacing increases. The HPBWs for those beams are depicted in Fig. \ref{c} for variable element spacing in UCCA arrays on both the azimuthal and the elevation plane. Here, it can be seen that as the spacing increases, the HPBW decreases in a UCCA array. Of course, these narrower beams come at the cost of larger antenna area sizes; see Fig. \ref{d}. Here, please notice that due to the use of mmWave frequency, the array size does not grow significantly, although it increases with the square of the antenna element spacing.  

In the following numerical examples, we consider UCCA array with two different element spacings, 1) $d_i=2\lambda$, and 2) $d_i=3\lambda$, namely $\rm{UCCA}_{2\lambda}$, and $\rm{UCCA}_{3\lambda}$, respectively. Now we find the beam packing gain $G_{\rm{BP}}$ of the above two arrays w.r.t. the conventional RPA array ($d_x=d_y=\lambda/2$). The HPBW of $\rm{UCCA}_{2\lambda}$: $(\theta_{\rm{3dB}},\phi_{\rm{3dB}})=(3.1^\circ,5.4^\circ)$, and that of $\rm{UCCA}_{3\lambda}$: $(\theta_{\rm{3dB}},\phi_{\rm{3dB}})=(2^\circ,3.6^\circ)$. As for the RPA array, these beamwidths are $(\theta_{\rm{3dB}},\phi_{\rm{3dB}})=(12^\circ,20.7^\circ)$.  Using (\ref{BP}), we obtain the beam packing gain of $\rm{UCCA_{2\lambda}}$, and $\rm{UCCA_{3\lambda}}$  arrays over the RPA array as 9 and 30 times, respectively.

In the next numerical example, our goal is to observe the minimum spatial separation distance as a function of the base station (BS) transmitter range using (\ref{ds1}), and (\ref{ds2}); see Fig. {\ref{Fig12}} and {\ref{Fig121}}, where we use the practical BS range of 5G networks. Here, it can be noticed that as expected the $\rm{UCCA_{3\lambda}}$, performs better than the $\rm{UCCA_{2\lambda}}$ array in placing the UEs more closely in a ultra dense network. For instance, at a distance 50 m from the BS the $\rm{UCCA}_{2\lambda}$, and $\rm{UCCA}_{3\lambda}$ are capable of offering 3.87 and 5.81 times more closely spaced UEs w.r.t. the RPA array while considering $S_{\phi_{3\rm{dB}}}$, and for $S_{\theta_{3\rm{dB}}}$ these numbers are 2.42 and 3.47 times, respectively. 

 To get a better insight into the leakage power, we list the maximum SLL of the above arrays in Table. \ref{sll} for performance comparison.
\begin{table}{}
\begin{center}
\caption{Side Lobe Level}
\begin{tabular}{|c|c|c|c|}
\hline
\textbf{SLL\,(dB)}&{RPA}& {$\rm{UCCA}_{2\lambda}$}&{$\rm{UCCA}_{3\lambda}$}\\
\hline
\textbf{Maximum $\rm{SLL}_\theta$} &-24.11&-18.06&-15.64 \\
 \hline 
\textbf{Maximum $\rm{SLL}_\phi$} &-14.64&-14.07&-11.55\\
 \hline 
\end{tabular}
\label{sll}
\end{center}
\end{table}
It can be seen that as expected the maximum SLL is higher in $\rm{UCCA}$ array than RPA array and among two UCCA arrays the $\rm{UCCA_{3\lambda}}$ is expected to suffer most in terms of inter-user interference. Notably, for all three arrays, this interference is more significant on the azimuthal plane than on the elevation plane.

\begin{table}{}

\begin{center}
\caption{Performance on the elevation plane}
\begin{tabular}{|c|c|c|}
\hline
\textbf{Array}&\textbf{SINR}\,(dB)& \textbf{{$\rm{\mathbf{SE}}$}}\,(bps/Hz)\\
\hline
RPA & 6.03  & 2.32 \\
\hline
 $\rm{UCCA}_{2\lambda}$ & 12.09 & 4.10 \\
 \hline 
$\rm{UCCA}_{3\lambda}$& 12.52 & 4.24\\
 \hline 
\end{tabular}
\label{tt1}
\end{center}
\begin{center}
\caption{Performance on the azimuthal plane}
\begin{tabular}{|c|c|c|}
\hline
\textbf{Array}&\textbf{SINR}\,(dB)& \textbf{{$\rm{\mathbf{SE}}$}}\,(bps/Hz)\\
\hline
RPA & -1.37  & 0.79 \\
\hline
$\rm{UCCA}_{2\lambda}$ & 10.88 & 3.73 \\
 \hline 
$\rm{UCCA}_{3\lambda}$& 12.15 & 4.12\\
 \hline 
\end{tabular}
\label{t1}
\end{center}
\end{table}

 Finally, we want to compare the SINR, and maximum spectral efficiency offered by the arrays of interests. We consider all the UEs are distributed within $0^{\circ}$ to $90^{\circ}$ in both the planes (azimuthal and elevation) and the signal of interest is at $(\theta_0,\phi_0)=(30^{\circ},60^{\circ})$. We perform 10000 realizations with arbitrary angles for $U=10$ interfering UEs on the boundary. We assume that the antenna elements are isotropic, there is no inter-element coupling, and the inner product of channel vectors is very much related to the antenna array pattern in far-field approximation \cite{pratschner2019large}. For user equal power allocation, and SNR = $10$ dB, the SINR, and maximum spectral efficiency for the arrays are calculated using (\ref{se}), and presented in Table \ref{t1}, and \ref{tt1} for comparison. Notice that the UCCA achieves significant improvement on SINR, and SE than the RPA array. $\rm{UCCA}_{2\lambda}$, and $\rm{UCCA}_{3\lambda}$ arrays are capable of providing 4.72 times and  5.22 times more spectral efficiency than the RPA array on the azimuthal plane. As expected, the $\rm{UCCA}_{3\lambda}$  shows slightly better performance than the $\rm{UCCA}_{2\lambda}$ due to its narrower HPBW. On the elevation plane, the $\rm{UCCA}_{2\lambda}$, and $\rm{UCCA}_{3\lambda}$ provide more than 6 dB gain in SINR while obtaining 1.77 and 1.83 times spectral efficiency gain, respectively over the RPA array.
 
\section{Conclusions}
In this paper, a comprehensive performance analysis of large aperture UCCA arrays was presented considering the conventional RPA array as the baseline performer. Our numerical study demonstrated that in a user dense 5G network, a larger aperture UCCA array performs better than the RPA array due to its capability of generating narrower beams while operating with a lesser number of antenna elements. For instance, a UCCA array with 98 antenna elements and $2\lambda$ antenna element spacing offered approximately 9 times beam packing gain, 4.72, and 1.77 times spectral efficiency gain on the azimuthal and the elevation plane, respectively over a $ 10 \times 10$ RPA array. Those gains were enhanced to 30, 5.22, and 1.83, respectively when the antenna element spacing was increased to  $3\lambda$.

\bibliographystyle{IEEEtran}
\bibliography{Reference.bib}

\begin{thebibliography}{1}
\providecommand{\url}[1]{#1}
\csname url@samestyle\endcsname
\providecommand{\newblock}{\relax}
\providecommand{\bibinfo}[2]{#2}
\providecommand{\BIBentrySTDinterwordspacing}{\spaceskip=0pt\relax}
\providecommand{\BIBentryALTinterwordstretchfactor}{4}
\providecommand{\BIBentryALTinterwordspacing}{\spaceskip=\fontdimen2\font plus
\BIBentryALTinterwordstretchfactor\fontdimen3\font minus
  \fontdimen4\font\relax}
\providecommand{\BIBforeignlanguage}[2]{{%
\expandafter\ifx\csname l@#1\endcsname\relax
\typeout{** WARNING: IEEEtran.bst: No hyphenation pattern has been}%
\typeout{** loaded for the language `#1'. Using the pattern for}%
\typeout{** the default language instead.}%
\else
\language=\csname l@#1\endcsname
\fi
#2}}
\providecommand{\BIBdecl}{\relax}
\BIBdecl

\bibitem{af}
E.~Bjornson, L.~Van~der Perre, S.~Buzzi, and E.~G. Larsson, ``Massive mimo in
  sub-6 ghz and mmwave: Physical, practical, and use-case differences,''
  \emph{IEEE Wireless Communications}, vol.~26, no.~2, pp. 100--108, 2019.

\bibitem{rpa}
F.~B{\o}hagen, P.~Orten, and G.~{\O}ien, ``Optimal design of uniform
  rectangular antenna arrays for strong line-of-sight mimo channels,''
  \emph{EURASIP Journal on Wireless Communications and Networking}, vol. 2007,
  pp. 1--10, 2007.

\bibitem{rr2}
I.~Zakia, ``Maximizing the sum rate of massive mimo with rectangular planar
  array and mrt beamforming,'' in \emph{2019 IEEE 89th Vehicular Technology
  Conference (VTC2019-Spring)}.\hskip 1em plus 0.5em minus 0.4em\relax IEEE,
  2019, pp. 1--5.

\bibitem{a03}
M.~I. Dessouky, H.~Sharshar, and Y.~Albagory, ``Efficient sidelobe reduction
  technique for small-sized concentric circular arrays,'' \emph{Progress In
  Electromagnetics Research}, vol.~65, pp. 187--200, 2006.

\bibitem{pratschner2019large}
S.~Pratschner, D.~L{\"o}schenbrand, S.~Schwarz, T.~Zemen, and M.~Rupp, ``Large
  aperture antenna array design for cellular los massive mimo,'' in \emph{2019
  53rd Asilomar Conference on Signals, Systems, and Computers}.\hskip 1em plus
  0.5em minus 0.4em\relax IEEE, 2019, pp. 1404--1408.

\bibitem{b2}
B.~Hamdi, S.~Limam, and T.~Aguili, ``Uniform and concentric circular antenna
  arrays synthesis for smart antenna systems using artificial neural network
  algorithm,'' \emph{Progress In Electromagnetics Research B}, vol.~67, pp.
  91--105, 2016.

\bibitem{b5}
G.~Fokin, S.~Bachevsky, and V.~Sevidov, ``System level performance evaluation
  of location aware beamforming in 5g ultra-dense networks,'' in \emph{2020
  IEEE International Conference on Electrical Engineering and Photonics
  (EExPolytech)}.\hskip 1em plus 0.5em minus 0.4em\relax IEEE, 2020, pp.
  94--97.

\bibitem{b3}
J.-J. Park, B.-J. Bae, and T.-G. Chang, ``Investigation of the sinr behavior of
  the beamforming-applied cellular cdma base station receiver,'' in \emph{WCNC.
  1999 IEEE Wireless Communications and Networking Conference (Cat. No.
  99TH8466)}, vol.~2.\hskip 1em plus 0.5em minus 0.4em\relax IEEE, 1999, pp.
  688--692.

\bibitem{b1}
W.~Li, X.~Huang, and H.~Leung, ``Performance evaluation of digital beamforming
  strategies for satellite communications,'' \emph{IEEE Transactions on
  Aerospace and Electronic systems}, vol.~40, no.~1, pp. 12--26, 2004.

\end{thebibliography}
\end{document}